\documentclass{article}
\usepackage[utf8]{inputenc}
\usepackage{amsmath}
\usepackage{graphicx}
\usepackage{subfig}
\usepackage{authblk}
\usepackage[dvipsnames]{xcolor}
\usepackage[margin=1in]{geometry}
\usepackage{multirow}
\usepackage{enumerate}
\usepackage{titlesec}

\usepackage{csquotes}
\usepackage[english]{babel}

\usepackage[style=numeric,sorting=none ]{biblatex}
\newcommand{\authorcite}[1]{\citeauthor{#1}\,\supercite{#1}}
\newtheorem{assumption}{Assumption}

\titleformat*{\subsection}{\large\itshape}

\title{Reducing bias in difference-in-differences models using entropy balancing}
\author[1]{Matthew Cefalu}
\author[1]{Brian G. Vegetabile}
\author[1]{Michael Dworsky}
\author[1]{Christine Eibner}
\author[1]{Federico Girosi}
\affil[1]{RAND Corporation}

\date{\today}

\begin{document}

\maketitle



\begin{abstract}

\textbf{Objective.} To illustrate the use of entropy balancing in difference-in-differences analyses when pre-intervention outcome trends suggest a possible violation of the parallel trends assumption.

\textbf{Data sources.} Simulated data and Medicare Advantage encounter data from 2014 to 2017 was used.

\textbf{Study Design.} We describe a set of assumptions under which weighting to balance intervention and comparison groups on pre-intervention outcome trends leads to unbiased difference-in-differences estimates even when pre-intervention trends in group mean outcomes are not parallel. We provide a simulation study highlighting the practical implications, and we apply the methodology to an evaluation of the Medicare Advantage Value-Based Insurance Design Model test. 

\textbf{Data Collection/Extraction methods}. Simulation analyses conducted in R and Medicare Advantage encounter data accessed through the Integrated Data Repository.

\textbf{Principal Findings.} Entropy balancing of pre-intervention outcomes trends requires two key assumptions to identify the average treatment effect on the treated: (1) persistence of the similarity of outcome trends from before to after the intervention; and (2) overlap between intervention and comparison groups in the pre-intervention outcome trends. Simulated results verify that entropy balancing of pre-intervention outcomes trends can remove bias even when unweighted pre-intervention outcome trends are not parallel between intervention and comparison groups. In some scenarios, entropy balancing appears to perform better than matching on pre-intervention trends. In our empirical example, MA VBID was associated with a 0.16 (0.11--0.21) and 0.14 (0.04--0.25) increase in the number of primary care and specialty care visits, respectively. Entropy balancing on pre-intervention outcome trends resulted in meaningfully different effects of VBID on specialty care visits compared to the unweighted difference-in-differences estimates.

\textbf{Conclusions.} Entropy balancing of pre-intervention outcomes trends can reduce bias in difference-in-differences analyses when the parallel trends assumption is not directly satisfied, and thus may enable researchers to use difference-in-differences designs in a wider range of observational settings than previously acknowledged.

\end{abstract}



\section{Introduction}

Assessing the impact of health policies frequently requires causal inference using observational research designs because randomized controlled trials are infeasible. When both pre- and post-intervention outcomes for the intervention and comparison groups are available, difference-in-differences (DD) is a widely used research design for causal inference.\supercite{wherry2016early,hanchate2015massjoint,osborne2015nsqip,lasser2014massreadmit,smulowitz2011massED,mcwilliams2016aco,mcwilliams2013aco} DD uses the difference between intervention and comparison groups in pre-intervention outcomes to control for permanent differences across groups in unobservable factors that affect outcomes; changes in the difference in group means after the intervention are attributed to the intervention.\supercite{wooldridge2002econometric} DD delivers unbiased estimates of causal effects under the assumption that the intervention and comparison groups would experience identical changes in post-intervention outcomes over time in the absence of the intervention, an assumption commonly referred to as \textit{parallel trends}.\supercite{angrist2008mostly} 

Although it is possible to estimate intervention effects using DD with data on group-level mean outcomes, it is common for researchers to carry out a DD study by estimating a regression model on individual-level microdata, sometimes referred to as the \emph{microlevel} DD estimator.\supercite{daw2018matching} Among other advantages, the use of microdata allows researchers to control for individual-level covariates, making the DD design more robust to observable changes in group composition. In the microlevel DD, the parallel trends assumption on unconditional group means can be replaced with a weaker assumption that post-intervention trends in group outcomes conditional on included covariates would be parallel in the absence of the intervention.\supercite{angrist2008mostly, stuart2014using} Regardless of the exact form, the parallel trends assumption is inherently untestable since the outcomes for the intervention group absent the intervention can never be observed.

Recently, researchers have combined matching with microlevel DD to further improve balance between groups on baseline observables, potentially including matching on pre-intervention outcomes.\supercite{stuart2014using} Whether matching on pre-intervention trends or other baseline observables actually reduces the bias of DD estimators is subject to debate. Prior simulation studies have shown that matching combined with a DD analysis can lead to bias reduction when compared to the usual DD analysis.\supercite{lindner2019difference,ryan2015we} Others, meanwhile, have argued that matching on pre-intervention outcome levels is insufficient to eliminate bias.\supercite{chabe2015analysis} In a recent paper, \authorcite{daw2018matching} point out that naively matching time-varying covariates or pre-intervention outcome levels can actually \emph{increase} bias. \authorcite{daw2018matching} also considered matching on pre-intervention trends and found that matching on pre-period trends failed to eliminate bias. They recommend that researchers should not use a DD study design when pre-intervention trends between the intervention and comparison groups are not parallel. In related work, \authorcite{arkhangelsky2019synthetic} propose a synthetic DD approach that weights comparison subjects based on pre-intervention outcome levels.

In this paper, we reexamine the questions of whether, and how, to combine DD with matching, weighting, or other approaches designed to balance pre-intervention outcome trends and characteristics. We consider settings where researchers have microdata covering multiple pre-intervention time periods, making it possible to reweight the comparison group based on information about individual-level \emph{trends} in pre-intervention outcomes. We discuss assumptions under which weighting can remove bias relative to unweighted DD and describe a set of sufficient conditions for such estimators to yield unbiased estimates of causal effects. As an alternative to matching, we show how to use a weighting method known as \emph{entropy balancing} to obtain balance on pre-intervention outcome trends.\supercite{hainmueller2012entropy} Using a simulation analysis, we demonstrate that it is possible to obtain unbiased DD estimates of causal effects even when the parallel trends assumption is not satisfied by the unweighted group means. 

We present an empirical example drawn from a recently completed policy evaluation to illustrate the use of entropy balancing in a DD study. Under the Value-Based Insurance Design (VBID) Model Test, insurers offering Medicare Advantage (MA) coverage (Medicare Advantage Organizations, or \textit{MAOs}) in certain states were allowed to modify benefit design based on a patient's health status, thereby encouraging patients with specified chronic conditions to increase utilization of high-value care and avoid costly and harmful complications. Using MA encounter data from 2014 to 2017, we estimate the impact of MA VBID on the number of primary care visits and the number of specialty care visits among VBID-eligible beneficiaries.



\section{Methods \label{sec:methods}}



\subsection*{Defining the Effect}

We define the causal effect of interest using potential outcomes in continuous time.\supercite{rubin2005causal} Let $A_i$ denote a binary indicator that the $i$-th individual was subject to the intervention. Further, assume that outcome data is available for all times $t \geq 0$, and the intervention is applied at time $t=t_e$, with $t_e > 0 $. We focus on \textit{instantaneous effects} at a fixed time point $t \geq t_e$, but our analysis also applies to \textit{time-averaged effects} defined over a range of post-intervention time points.

Let $Y_{i}(1,t)$ and $Y_{i}(0,t)$ denote the \textit{potential outcomes} that would be observed with and without the intervention, respectively, for the $i$-th individual at time $t$. We observe only one potential outcome at each time point $t$, which is denoted by the \textit{observed outcome} $Y_{i}(t) = A_i Y_{i}(1,t) + (1-A_i) Y_{i}(0,t)$.

For individual $i$, the instantaneous effect at time $t$ is defined as the difference in their potential outcomes,

\begin{align*}
    \tau_i(t) = Y_{i}(1,t) - Y_{i}(0,t).
\end{align*}

\noindent Averaging these individual-level instantaneous effects over the population that was subject to the intervention defines the estimand of interest, the \emph{average treatment effect on the treated} (ATT) at time $t$,

\begin{align*}
    ATT(t) &= E[\tau(t) | A=1].
\end{align*}


\subsection*{Difference-in-Differences}

Microlevel DD can be formulated in many different ways, including through differences in groups means. For presentation, we define DD models as regression models. A microlevel DD model that specifies common time effects as a polynomial of order $P$ can be written as follows:



\begin{align}
    y_{it} = \alpha + \sum_{p=1}^P \beta_p t^p + \beta^A A_i + \tau A_i E_t + \varepsilon_{it} \label{eqn:DDpoly}
\end{align}

where $E_t$ is an indicator for the post-intervention time period ($t \geq t_e$), $\alpha$ is a constant, and $\varepsilon_{it}$ is an error term. Alternatively, a microlevel DD model with nonparametric time effects can be written as follows:

\begin{align}
    y_{it} = \mu_t + \beta^A A_i + \tau A_i E_t + \varepsilon_{it} \label{eqn:DDnonparm}
\end{align}

where $\mu_t$ denotes fixed effects for each time period and $\varepsilon_{it}$ is an error term.

In our simulation analysis, we consider DD models that control for time linearly (Equation \ref{eqn:DDpoly} with $P=1$), quadratically (Equation \ref{eqn:DDpoly} with $P=2$), or nonparametrically (Equation \ref{eqn:DDnonparm}). However, the theoretical arguments in the remainder of this section are not tied to any one specification of DD. 

\subsection*{Identifying assumptions}

The ATT is a function of unobservable potential outcomes, and is not identified from observed data without additional assumptions. The key identifying assumption for DD is the parallel trends assumption. 
To formalize this assumption in the context of weighted DD, we define $\delta_{i}(a,t) \equiv \frac{\partial}{\partial t} Y_i(a,t)$ as the potential outcome trend for individual $i$ under intervention assignment $a$ at time $t$. $\delta_{i}(0,t)$ is the potential outcome trend absent the intervention, and $\delta_{i}(1,t)$ is the potential outcome trend under intervention. The parallel trends assumption for weighted DD is as follows: 

\begin{assumption}[Parallel Trends] \label{assump:parallel}
There exist weights $w$ such that, 
\begin{align*} 
    E[\delta(0,t) | A=1] = E[ w \delta(0,t) | A=0] \mbox{ for all } t \geq t_e.
\end{align*}
\end{assumption}

\noindent Assumption \ref{assump:parallel} states that it is possible to find a sub-population of the comparison group that, after weighting, exhibits identical post-intervention potential outcome trends to those in the intervention group absent the intervention. In the Appendix, we prove that Assumption \ref{assump:parallel} is sufficient to identify the average treatment effect on the treated using DD. 

The parallel trends assumption is untestable because it depends on unobserved potential outcomes. It also provides no indication of how to find $w_i$. To guide the derivation of $w_i$ in practice, we introduce two alternative assumptions involving pre-intervention outcome trends that, together, imply Assumption \ref{assump:parallel}.

\begin{assumption}[Parallel Pre-Intervention Observed Outcome Trends] \label{assumpt:observed}
There exist weights $w$ such that, 
\begin{align*} 
    E[\delta(t) | A=1] = E[ w \delta(t) | A=0] \mbox{ for all } t < t_e.
\end{align*}
\end{assumption}
    
\noindent Assumption \ref{assumpt:observed} states that, among the comparison group, it is possible to find a weighted sub-population that exhibits similar pre-intervention outcome trends to the intervention group. Unlike Assumption 1, this is a testable assumption, as it relates to observed outcomes only.

In order for Assumption \ref{assumpt:observed} to have any implications for the validity of DD, it is necessary to make a further assumption about the relationship between pre- and post-intervention potential outcomes.

\begin{assumption}[Persistence of Similarity of Potential Outcome Trends] \label{assump:persistence}
For any $w$, if
\begin{align*}
      E[\delta(0,t) | A = 1] = E[w\delta(0,t) | A = 0] \mbox{ for all } t < t_e 
\end{align*}
then,
\begin{align*}
    E[\delta(0,t) | A = 1] = E[w\delta(0,t) | A = 0] \mbox{ for all } t \geq t_e.
\end{align*}
\end{assumption}

\noindent Assumption \ref{assump:persistence} states that if the groups have parallel potential outcome trends in the pre-intervention period, then they also have parallel post-intervention potential outcome trends absent the intervention. Assumption \ref{assump:persistence} allows arbitrary changes in outcome trends after the intervention, provided that such changes would affect both groups absent the intervention, i.e., it allows for \textit{common shocks}. Assumption \ref{assump:persistence} formalizes the intuition that motivates the standard practice of testing for parallel pre-intervention trends in DD studies with multiple pre-intervention time periods. 


Together, Assumptions \ref{assumpt:observed} and \ref{assump:persistence} imply Assumption \ref{assump:parallel}. Assumptions \ref{assumpt:observed} and \ref{assump:persistence} are thus jointly sufficient for DD to identify the ATT (see Appendix). 
Furthermore, the sufficiency of Assumptions \ref{assumpt:observed} and \ref{assump:persistence} for the DD to estimate the ATT provides a path forward for deriving the weights $w_i$: $w_i$ should be chosen to balance pre-intervention outcome trends between the intervention and comparison groups. This can be done using existing approaches designed to balance pre-intervention characteristics, including matching,\supercite{daw2018matching} propensity score methods,\supercite{stuart2014using,linden2011applying} or entropy balancing. Below we discuss how to incorporate entropy balancing into weighted DD estimation. 

\subsection*{Entropy Balancing for Difference-in-Differences}

We describe the application of Hainmueller's\supercite{hainmueller2012entropy} \textit{entropy balancing} approach to calculating weights for DD that balance pre-intervention outcome trends, thereby satisfying Assumption \ref{assumpt:observed}. Methodological details are left to the Appendix. Entropy balancing selects weights that satisfy a series of \textit{balancing constraints} based on the distribution of the observables in intervention and comparison groups. Although previous studies have combined entropy balancing with DD, these applications have focused on balancing the levels of pre-intervention covariates rather than pre-intervention outcome trends.\supercite{marcus2013effect,parish2018using} The application of entropy balancing to pre-intervention trends in DD has not previously been analyzed in detail.

For DD, the most important variables to include in the balancing constraints are estimates of the pre-intervention outcome trends. We consider two approaches to estimating individual-level pre-intervention outcome trends for inclusion in the balancing constraints (see Appendix for details). The first strategy is to model the trends parametrically, for instance using linear regression to estimate a pre-intervention outcome trend for each individual. We expect this method will perform well when the individual-level trends are noisy, as the parametric trends will average out the noise over time. However, misspecification of the parametric trends used to define entropy-balancing weights may result in violation of the Assumption \ref{assumpt:observed} or \ref{assump:persistence} since the misspecified trends do not represent the true trends.


An alternative is to nonparametrically estimate the slope between each time point by taking first differences of the pre-intervention observed outcomes. We expect this method will perform well when observed outcome trends reflect heterogeneity across individuals, rather than transitory within-individual variability (or \emph{noise}) that weakens the relationship between pre-intervention observed outcome trends and post-intervention potential outcome trends. 

Regardless of how the pre-intervention outcome trends are estimated, Assumption \ref{assumpt:observed} relates to the true pre-intervention outcome trends while, in practice, we attempt to balance estimated pre-intervention outcome trends. If individual trends in outcomes reflect noise---perhaps due to measurement error or other transitory error components---then estimated outcome trends may be less \emph{reliable} (in the sense of Steiner et al.\supercite{steiner2011importance}) and we expect that our ability to remove bias will suffer. 

For entropy balancing to allow DD estimation of the ATT, one more assumption is required\supercite{hainmueller2012entropy}: the support of the intervention group's distribution of variables used in the balancing constraints must lie within the support for the comparison group, which we refer to as overlap. Overlap assumptions are standard in the literature on matching and related causal inference methods.\supercite{stuart2010matching} In our context, this assumption applies to the distribution of pre-intervention trends:

\begin{assumption}[Overlap in Pre-Intervention Trends] \label{assumpt:overlap}
\begin{align*} 
    \mbox{Pr}(A=1 | \delta_i(t)) < 1 ~~\forall~~ i : A_i=1
\end{align*}
\end{assumption}

Overlap may be violated if there are individuals in the intervention group with pre-intervention outcome trends that differ radically from the comparison group. In such cases, it may not be possible to find a set of weights that allows estimation of the ATT. In this case, it may still be possible to estimate a local average treatment effect defined on the subpopulation of the intervention group that satisfies overlap. 






\section{Simulation Study \label{sec:simulation}}
\subsection{Overview}

We demonstrate the use of weighted DD estimation using simulated data in order to evaluate the performance of alternative estimators and to highlight the assumptions necessary to reduce bias by balancing pre-intervention outcome trends. Technical details of these simulations are presented in the Appendix.

We focus primarily on two families of data generating processes (DGPs) that embed different assumptions about the distribution of post-intervention potential outcome trends in the intervention and comparison groups: (1) group mean trends are different with overlap in the individual-level trends; and (2) group mean trends are different without overlap in the individual-level trends. Figure \ref{fig:trends} presents histograms of individual-level trends that illustrate the difference between the scenarios with and without overlap. In these DGPs, outcomes are a linear combination of the intervention effect, a deterministic linear trend with heterogeneous slopes across individuals, and an error term. Individual-level trends in potential outcomes absent treatment are constant over time, which means that Assumption \ref{assump:persistence} is satisfied. Within each family of DGPs, we also explore the importance of reliability of the estimated pre-intervention outcome trends by simulating DGPs that vary in the degree of autocorrelation in the error term: DGPs with higher autocorrelation allow more reliable estimates of pre-intervention observed outcome trends.


Under each DGP, we compare the bias of three DD models that use different approaches for handling the differences in group mean pre-intervention outcome trends: an unadjusted approach, matching on the estimated linear trend, and entropy balancing on the estimated linear trend. For each of these approaches, we fit two DD models: one that includes time as a linear function (Equation \ref{eqn:DDpoly} with $P=1$) and one that includes time nonparametrically (Equation \ref{eqn:DDnonparm}). For matching, we use 1:1 matching without replacement and a caliper of 0.2 standard deviations. For entropy balancing, we balance on the first moments of the estimated trends. We report estimator performance in terms of the percent bias reduction relative to the bias of the unweighted DD estimator: an unbiased estimator will have 100\% bias reduction.

In results not shown here, we also specified a DGP such that the group counterfactual trends are the same. In this case, Assumption \ref{assump:parallel} is satisfied without weighting and DD is unbiased without weighting. All DD estimators are approximately unbiased for all values of the residual autocorrelation. This analysis simply confirms that the weighted DD estimators considered here do not introduce bias in situations when unweighted DD is also unbiased.

\subsection{Scenario 1: Overlap in the individual-level trends}



\begin{figure}
    \centering
    \includegraphics[width=\textwidth]{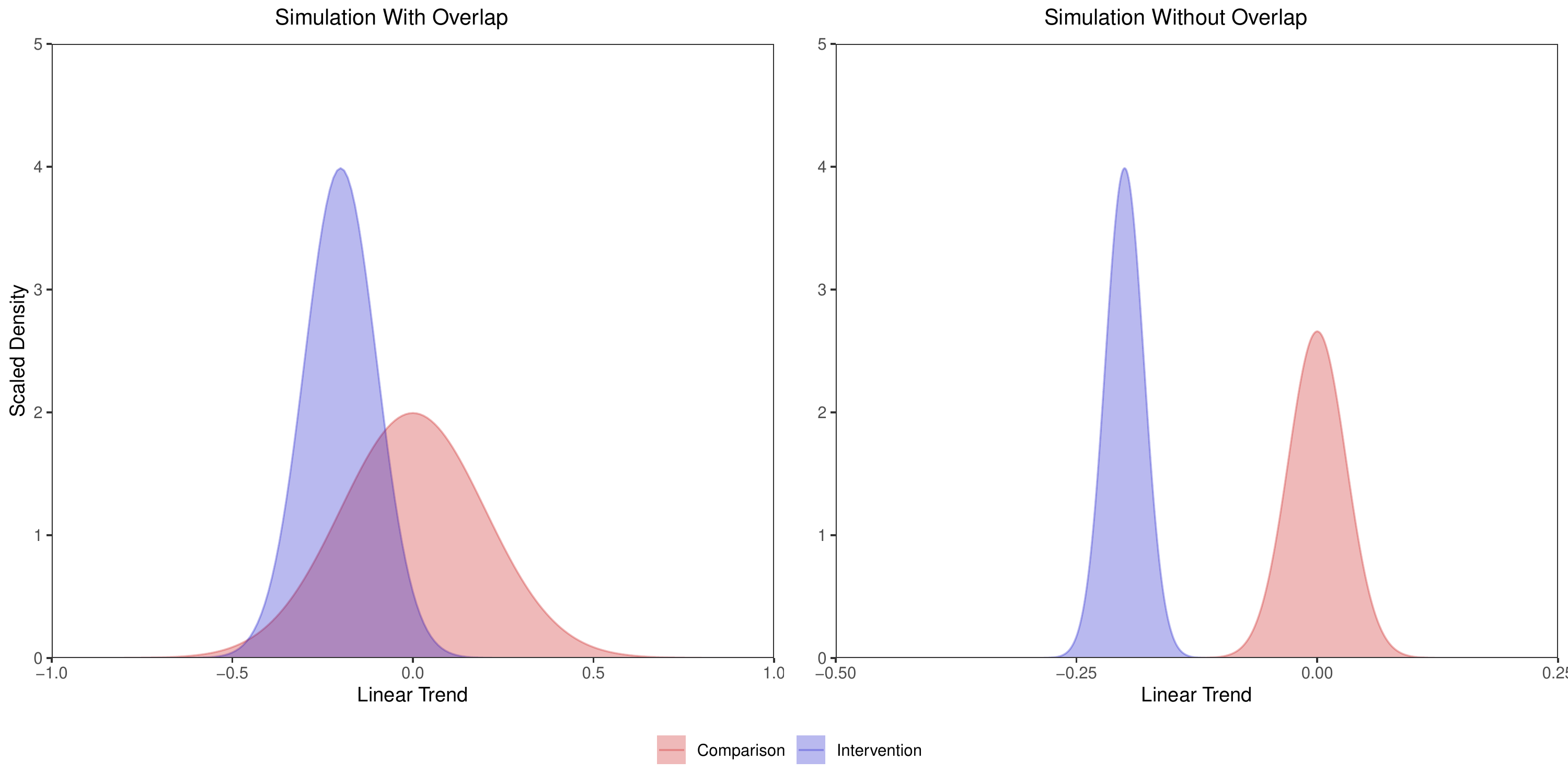}
    \caption{Distribution of individual-level linear time trends by intervention group.}
    \label{fig:trends}
\end{figure}

First, we simulate DGPs that are ideal for balancing pre-intervention trends: there are mean differences in the counterfactual outcome trends, but there is overlap in the individual-level trends between the intervention groups (satisfying Assumption \ref{assumpt:overlap}). The distribution of the slopes of the individual-level trends is illustrated in the left panel of Figure \ref{fig:trends}, which confirms that all values of the intervention group are represented in the comparison group.


Figure \ref{fig:fig1} reports the percent bias reduction comparing each estimator to the usual DD as a function of the residual autocorrelation, which varies from zero to 0.99. First, all four weighted DD estimators reduce bias. Second, given the balancing approach, estimators that control for linear trends (left panel) yield greater bias reduction than approaches that control for nonparametric time effects (right panel). This likely occurs because the true outcome trends are linear and are better approximated by the DD with time parameterized linearly. Third, for all estimators, the bias reduction increases as the residual autocorrelation increases. The pattern of improved bias reduction with increasing residual autocorrelation is likely to reflect the increasing reliability of the estimated outcome trends.\supercite{steiner2011importance} Fourth, entropy balancing on linear trends (left panel) removes most of the bias for all values of the residual autocorrelation. In contrast, a similar matching based estimator does not. 


\subsection{Scenario 2: No overlap in the individual-level trends}

\begin{figure}
    \centering
    \includegraphics[width=\textwidth]{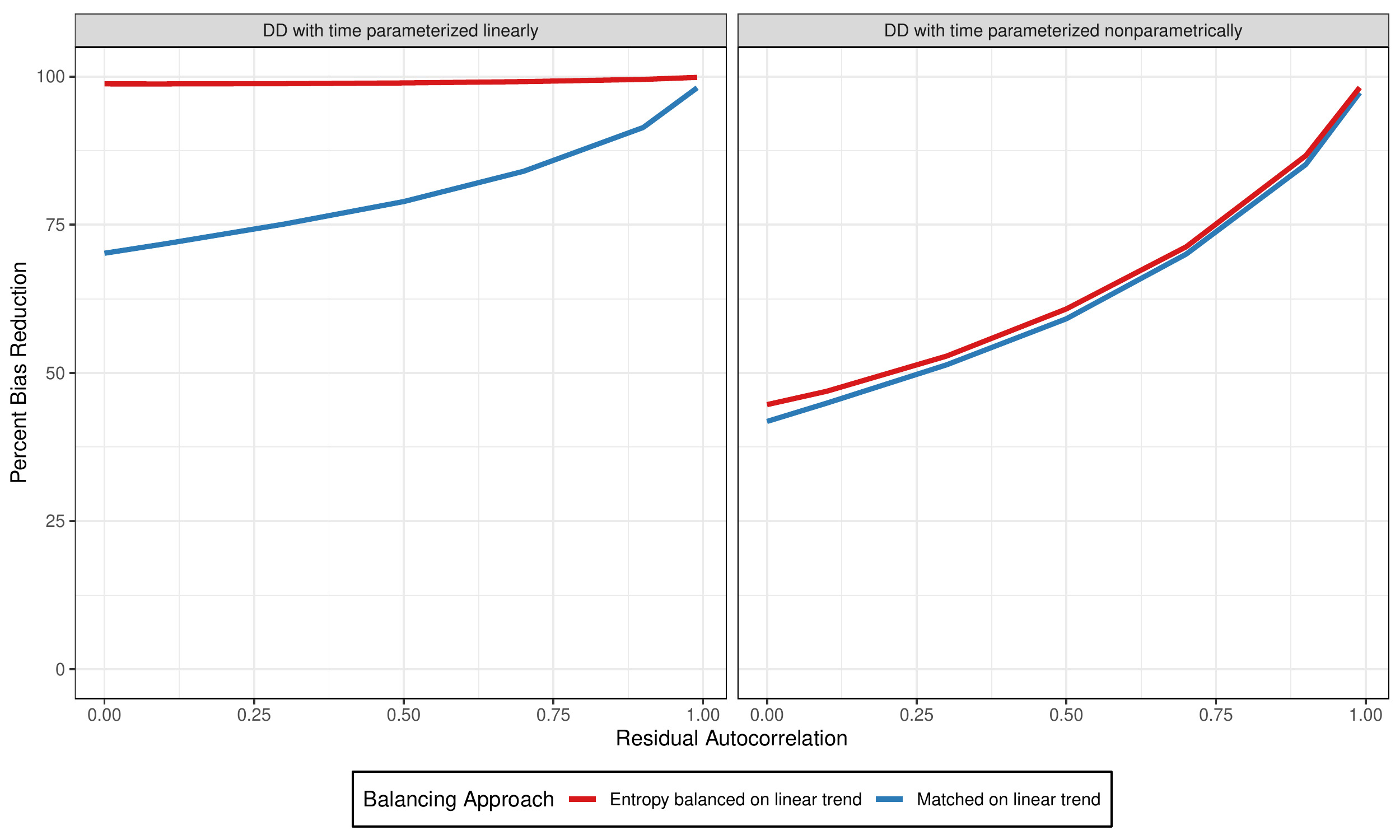}
    \caption{Percent bias reduction from the simulation study when group counterfactual trends are different with overlap in the individual-level trends. The left panel provides the results for DD with time parameterized linearly, and the right panel provides the results for DD with time parameterized nonparametrically.}
    \label{fig:fig1}
\end{figure}

\begin{figure}
    \centering
    \includegraphics[width=\textwidth]{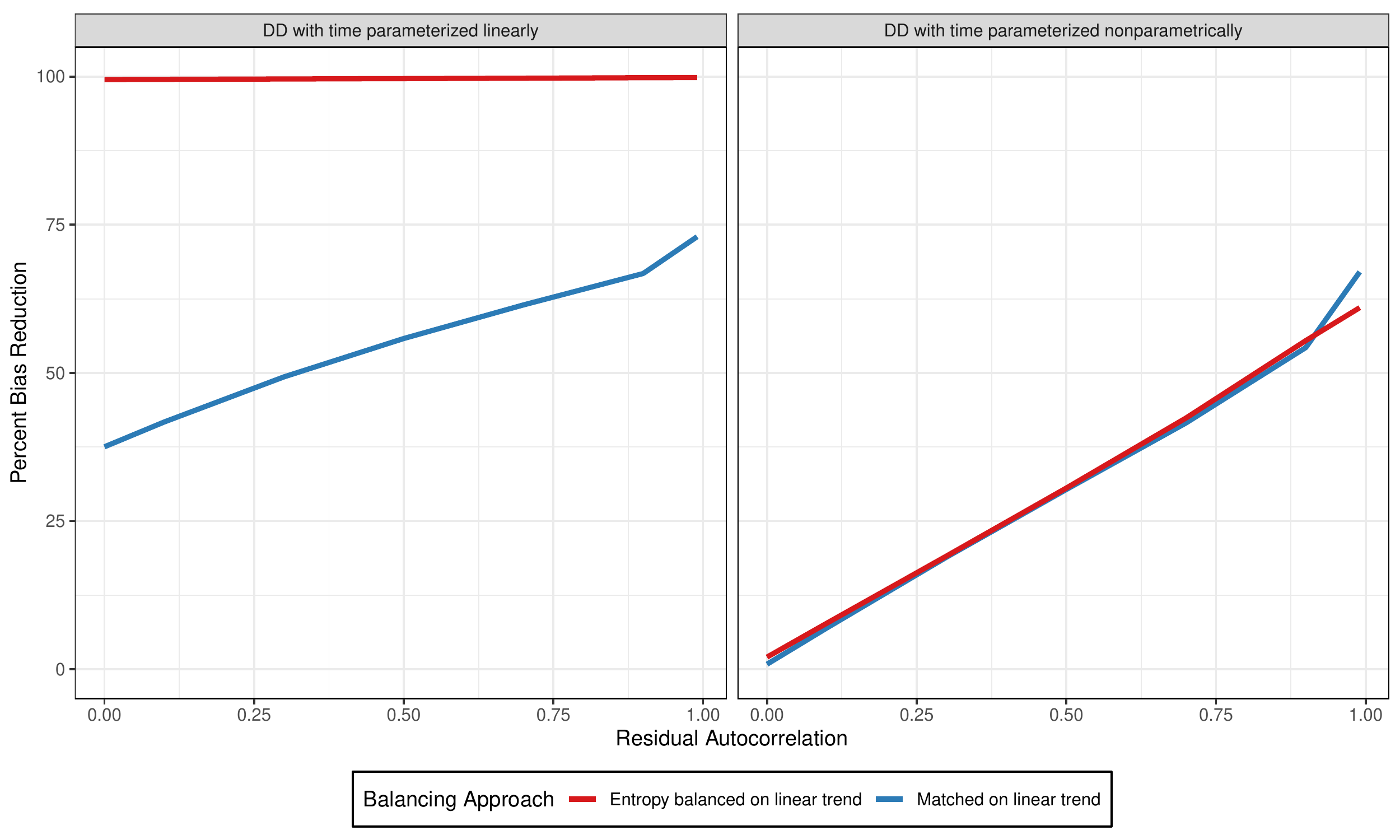}
    \caption{Percent bias reduction from the simulation study when group counterfactual trends are different without overlap in the individual-level trends. The left panel provides the results for DD with time parameterized linearly, and the right panel provides the results for DD with time parameterized nonparametrically.}
    \label{fig:fig2}
\end{figure}

Here, we simulate DGPs that violate the assumption of overlap in the individual-level trends (Assumption \ref{assumpt:overlap}). 
This is highlighted in the right panel of Figure \ref{fig:trends}, where we have plotted the distribution of individuals' linear time trends by intervention group. 
Figure \ref{fig:fig2} reports the percent bias reduction comparing each estimator to the usual DD. As in Figure \ref{fig:fig1}, bias reduction is plotted against the residual autocorrelation.




The general pattern of bias reduction is similar to that from the previous results, with two key distinctions. First, the balancing approaches do not remove all bias as the residual autocorrelation approaches 1. Second, the amount of bias reduction compared to the unadjusted model is less than that of the previous section. These results are due to the fact that there is no overlap in the trends, so the balancing approaches are unable to match the true pre-intervention trends even as the residual autocorrelation increases. 

An unexpected finding is that entropy balancing of the linear outcome trends combined with a DD model that parameterizes time linearly is approximately unbiased for all values of the autocorrelation---even without overlap. We believe this reflects the double robustness property of entropy balancing, which states (among other things) that entropy balancing augmented with a regression model for the outcome is consistent, as long as the outcome model is correctly specified.\supercite{qingyuan2016} Here, we have correctly formulated the outcome regression as a linear model, potentially satisfying the assumption needed for consistency. Technically, our DD model is misspecified: the group mean trend differs between intervention groups, but the model assumes a common trend. However, the double robustness property for such models may require only correct specification of the outcome model among the comparison group due to the estimand of interest being defined as the ATT. More work is needed to theoretically justify this result for DD models.  


\section{Impact of Medicare Advantage Value-Based Insurance Design \label{sec:VBID}}

To illustrate the use of entropy balancing in DD evaluation of a health policy intervention, we used data from the first and second annual evaluations of the MA VBID model test.\supercite{eibner2018first} VBID is an approach to health insurance design that tailors patient cost-sharing and other dimensions of benefit design on the basis of patients' health status. VBID aims to guide patients toward more appropriate utilization decisions, e.g., by reducing co-payments associated with health services or pharmaceuticals with a high clinical value given the patient's chronic conditions.\supercite{chernew2007value} Earlier demonstrations of VBID have shown promise in the employer-sponsored insurance market,\supercite{chernew2008impact,frank2012effect,choudhry2011full,choudhry2010pitney,choudhry2014five,gibson2011value,maciejewski2014value,yeung2017impact} but VBID had not previously been tested in Medicare beneficiaries aged 65 and over. 
Between 2017 and 2019, the Center for Medicare \& Medicaid Innovation (CMMI) within the Centers for Medicare \& Medicaid Services (CMS) conducted the first test of VBID in Medicare Advantage (MA), an intervention referred to as the MA VBID Model Test.\supercite{eibner2018first} 
In 2017, a total of 45 MA plans participated in the MA VBID model test. See \authorcite{eibner2018first} for details on the implementation, benefit designs, and impacts of MA VBID, as well as details on data sources and data construction for the sample presented here. 

In this paper, we provide DD estimates of the effect of VBID on per-beneficiary per year utilization of office-based primary care and specialty care. Table \ref{tab:descriptives} provides baseline characteristics and outcomes from 2016 of VBID-eligible beneficiaries in participating MA plans and their matched comparison beneficiaries in non-participating MA plans. Overall, baseline characteristics of beneficiaries are similar between VBID-participating and comparison MA plans, with beneficiaries in VBID plans being slightly older (77.8 vs 77.1) and having higher risk scores (1.80 vs 1.58). We apply entropy balancing using the full set of baseline characteristics from Table \ref{tab:descriptives} and the first-differences of the pre-intervention observed outcomes. After application of entropy balancing weights for both sets of outcomes, the means of the baseline characteristics from comparison beneficiaries match those of VBID beneficiaries exactly. Note that the outcome levels do not match, as the entropy balancing only seeks to balance pre-intervention outcome trends. 

\begin{figure}
    \centering
    \subfloat[]{\includegraphics[width=.45 \textwidth]{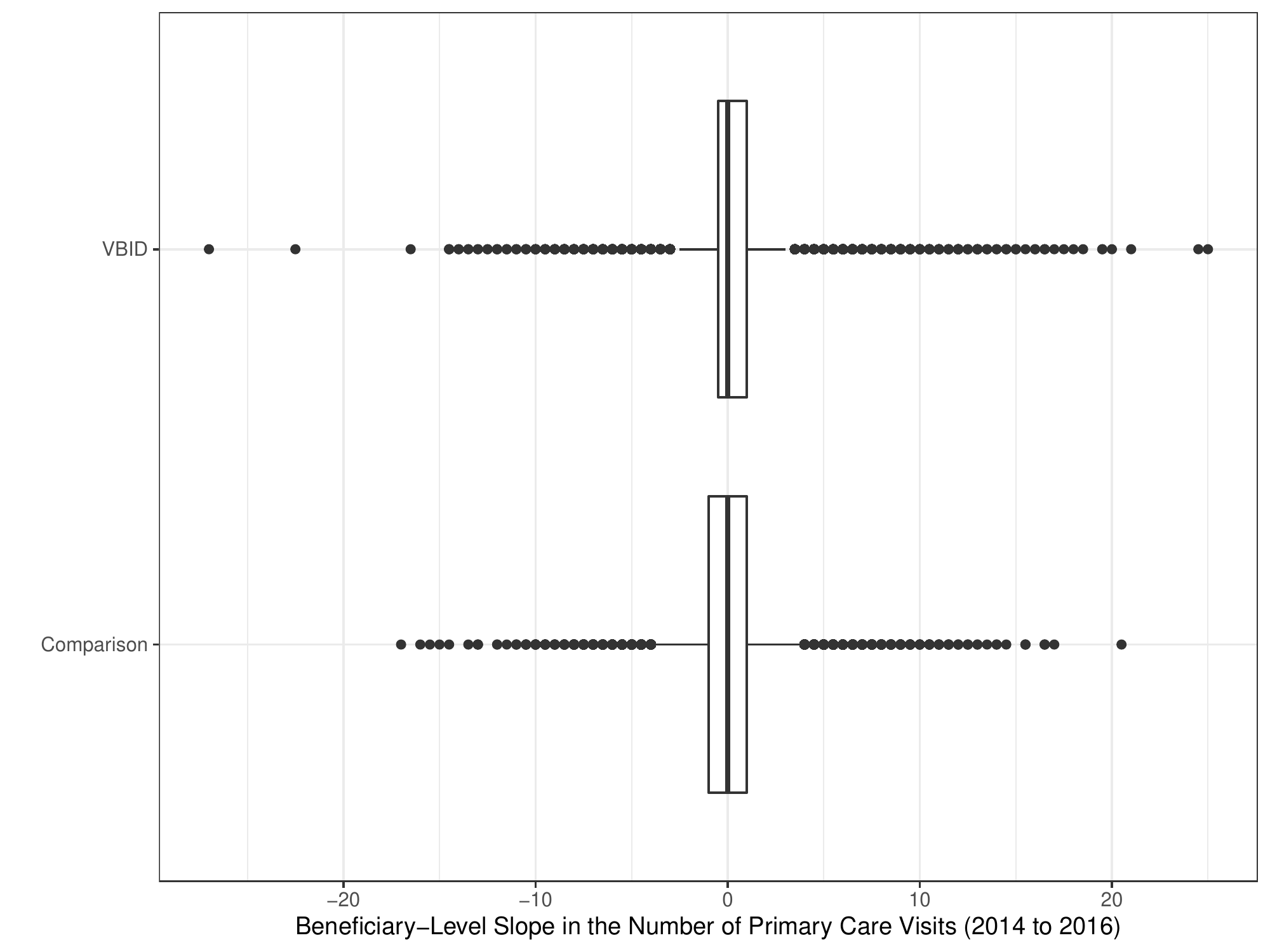}}
    \qquad
    \subfloat[]{\includegraphics[width=.45 \textwidth]{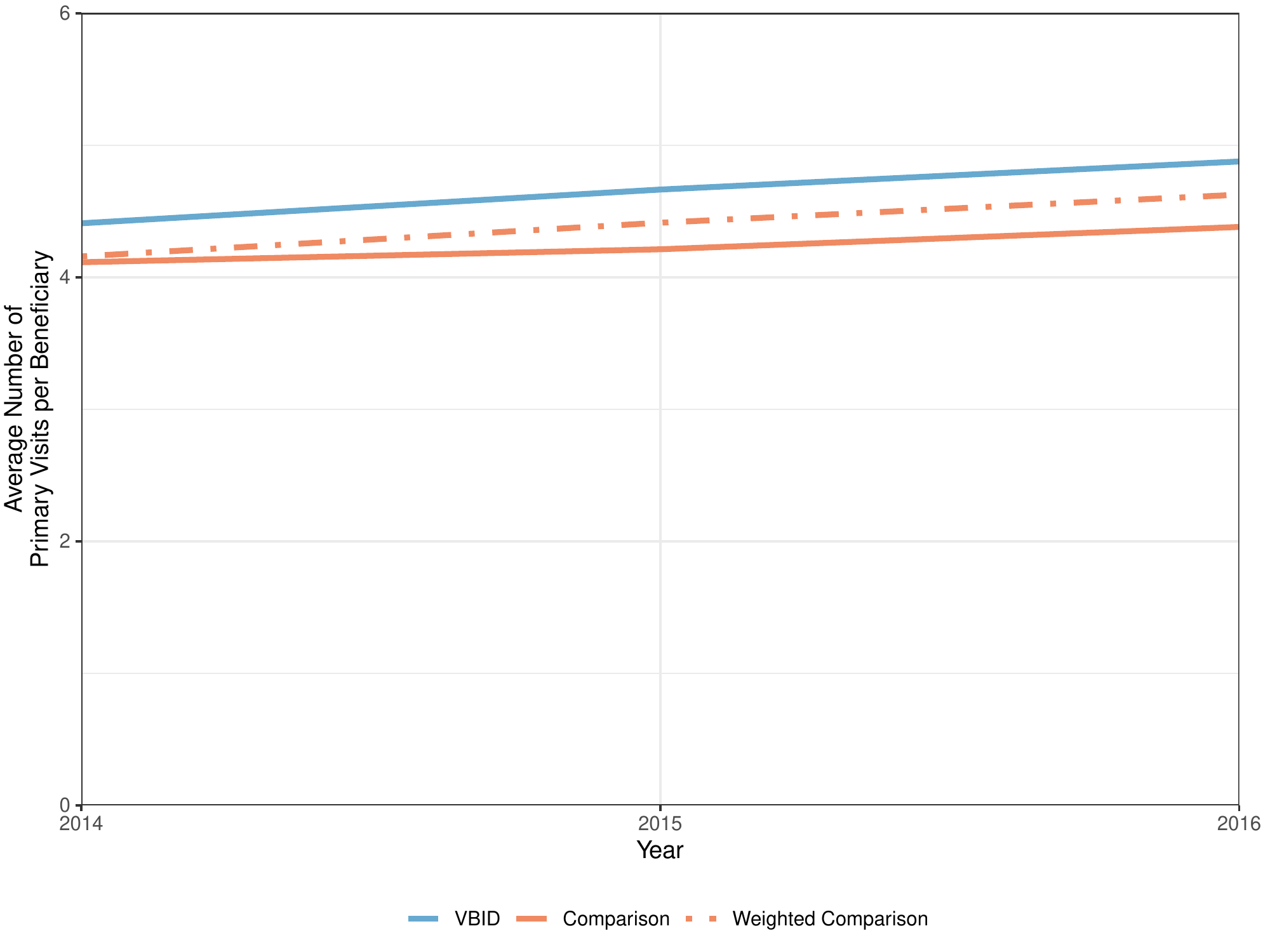}}
    \caption{(a) Overlap in the estimated beneficiary-level slope in the number of primary care visits, 2014-2016. (b) Average number of primary care visits from 2014 to 2017 for VBID-eligible beneficiaries in VBID-participating MA plans, comparison MA plans, and comparison MA plans after entropy balancing.}
    \label{fig:primary}
\end{figure}

\begin{figure}
    \centering
    \subfloat[]{\includegraphics[width=.45 \textwidth]{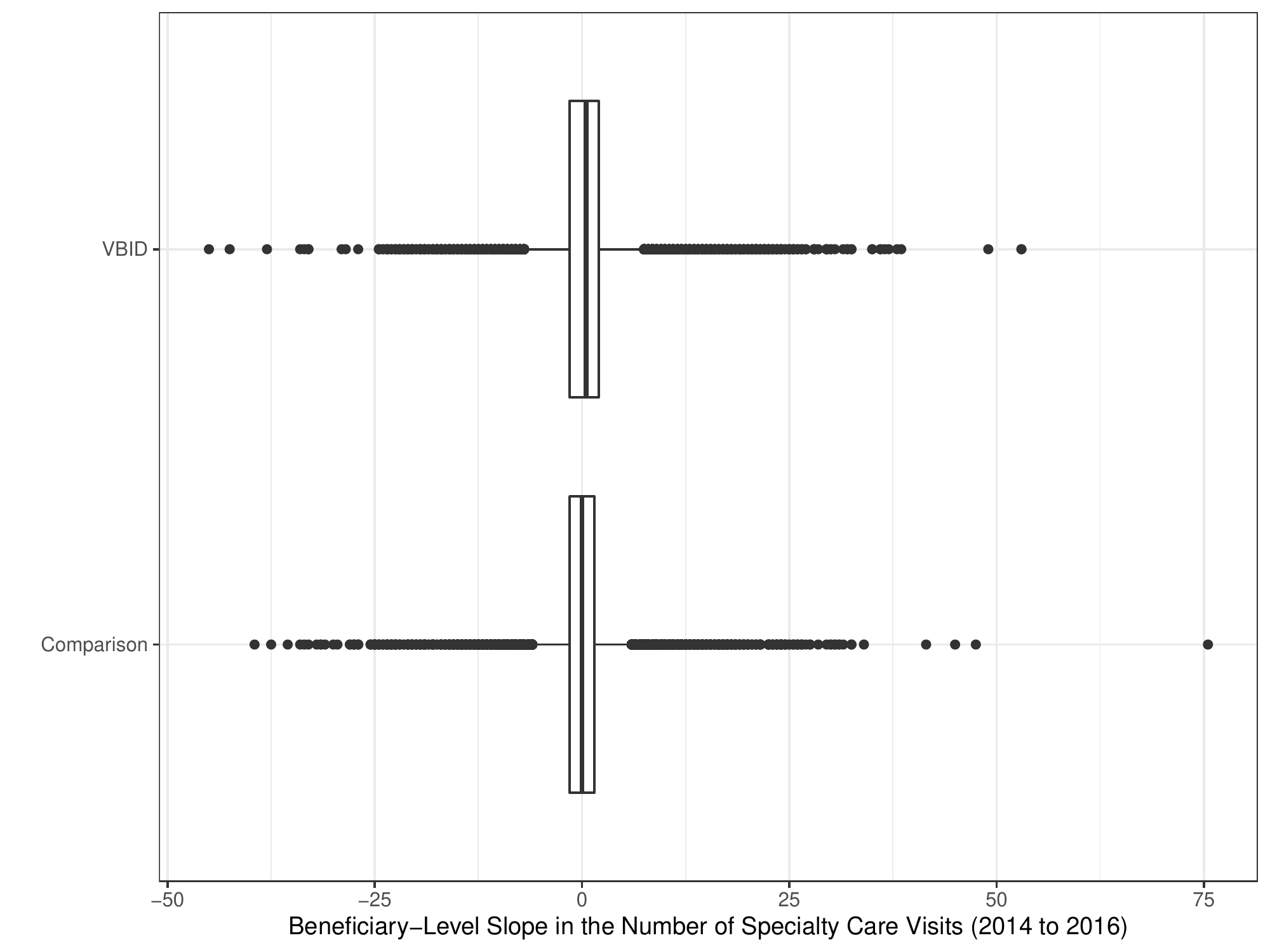}}
    \qquad
    \subfloat[]{\includegraphics[width=.45 \textwidth]{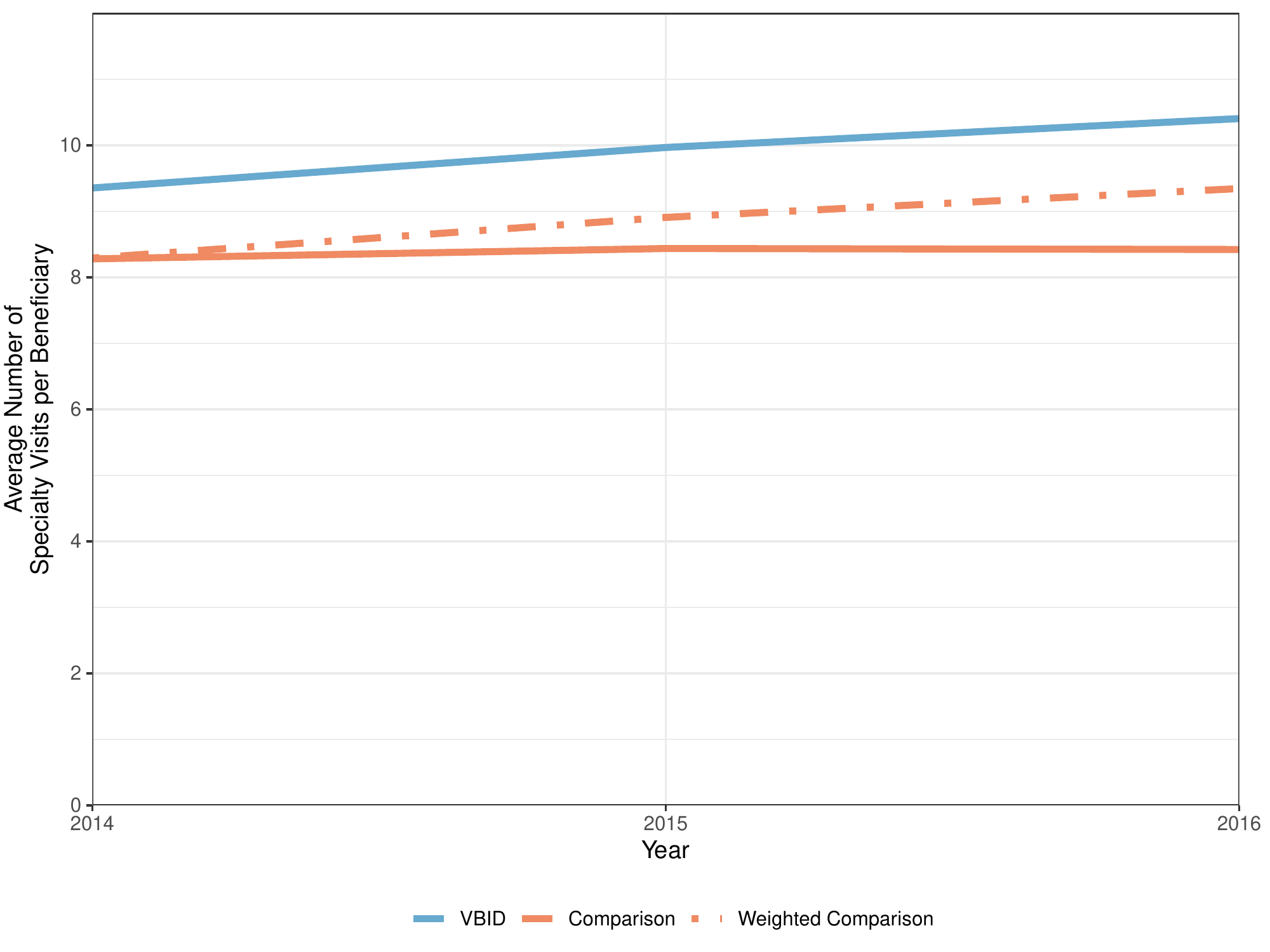}}
    \caption{(a) Overlap in the estimated beneficiary-level slope in the number of specialty care visits, 2014-2016. (b) Average number of specialty care visits from 2014 to 2017 for VBID-eligible beneficiaries in VBID-participating MA plans, comparison MA plans, and comparison MA plans after entropy balancing.}
    \label{fig:specialty}
\end{figure}

Figure \ref{fig:primary}(a) shows overlap between VBID and comparison groups in the beneficiary-level trend in the number of primary care visits from 2014 to 2016. Only a handful of the nearly 40,000 VBID-eligible beneficiaries lie outside the range of comparison beneficaries, suggesting sufficient overlap. Figure \ref{fig:primary}(b) shows the average number of primary care visits for VBID-eligible beneficiaries in VBID-participating plans and for comparison beneficiaries before and after entropy balancing. Trends from 2014 to 2016 were similar prior to entropy weighting, but comparison beneficiaries' utilization is increasing at a slower rate than VBID beneficiaries. A test of a departure of parallel trends prior to 2017 rejects the null hypothesis of parallel trends before entropy balancing ($p<0.001$), but not after ($p\approx 1$). This result is visually verified in Figure \ref{fig:primary}(b). Figures \ref{fig:specialty}(a) and \ref{fig:specialty}(b) show the same information for specialty visits, and similar conclusions are formed. A test of a departure of parallel trends prior to 2017 for specialty visits rejects the null hypothesis of parallel trends before weighting ($p<0.001$), but not after ($p\approx 1$). This result is visually verified in Figure \ref{fig:specialty}(b). 


The estimated increase in the number of primary care visits per beneficiary that is associated with VBID is 0.12 (0.08--0.17) before entropy balancing and 0.16 (0.11--0.21) after entropy balancing. A larger difference is observed for specialty care visits, with the increase in the number of visits associated with VBID estimated as 0.05 (-0.04--0.14) prior to entropy balancing and 0.14 ( 0.04--0.25) after entropy balancing---nearly three times larger an effect. In this example, DD after entropy balancing could lead to a substantially different conclusion about policy impacts than the conclusion one might draw from the unweighted DD estimate.

\section{Discussion \label{sec:discussion}}

Our theoretical developments and simulation studies provide practical guidance on when it is beneficial to balance pre-intervention trends in the outcomes between intervention groups. Although we investigated only a few DGPs, we expect similar results to hold across many other settings. The simulation results presented in this paper were limited to cases where the counterfactual trends were linear, which guarantees that Assumption \ref{assump:persistence} is satisfied. In the Appendix, we relaxed this assumption by simulating data with nonlinearities, and find that using nonparametric time simultaneously in entropy balancing and the DD model yields greater bias reductions when compared to balancing the estimated linear trends.

Based on the results of this study, we can offer some recommendations for health policy researchers. If the parallel trends assumption of the DD model is expected to hold for the group means, then there is no need to balance pre-intervention trends. However, entropy balancing of pre-intervention trends does not introduce bias. If there is uncertainty about whether the parallel trends assumption is expected to hold, then entropy balancing of pre-intervention outcome trends reduces bias relative to the unadjusted estimator. We also note that our theoretical and simulated results focus on outcome trends, and none of our findings justify balancing pre-intervention outcome levels: in this we agree with \authorcite{daw2018matching}.

The amount of bias reduction that can be achieved using entropy balancing for DD, or any other similar approach, is limited by how well the individual-level outcome trends can be estimated. A highly reliable estimate of the individual-level trends will remove nearly all of the bias, while a completely unreliable estimate will remove no bias. The mechanism through which the estimated individual-level trends have higher reliability should not matter, and we expect any highly reliable estimated trend to remove considerable bias. Higher reliability is achieved as the autocorrelation increases, as the residual error decreases, as the strength of the time trends increases, as the individual-level heterogeneity of the time trends increases, or as the number of pre-intervention time points increases (if balancing a parametric outcome trend), among others. Adding additional individuals will not, in general, improve the reliability of the estimated outcome trends. Care should be taken to estimate the pre-intervention outcome trends in a manner that maximizes reliability without sacrificing robustness. 


The assumptions presented in this paper are sufficient (not necessary) conditions for weighted DD to eliminate bias. It is possible that they can be relaxed or modified. In some applications, the parallel trends assumptions might be narrowed to specific time periods to account for anticipatory effects or a wash-out period. We also expect similar results to hold for other causal estimands identified using DD, such as the time-averaged ATT.



Entropy balancing for DD also offers a practical advantage for large-scale policy evaluations, such as the evaluation of MA VBID described in this paper, that analyze many outcome measures. Researchers may have difficulty identifying a comparison group that directly satisfies the parallel trends assumption uniformly across all outcomes. As in the MA VBID evaluation, entropy balancing on pre-intervention outcome trends can be used to refine a comparison group that can be used to analyze many different outcomes while reducing the bias of DD estimates when the parallel trends assumption is violated for specific outcomes. For instance, in the example presented in Section \ref{sec:VBID}, pre-intervention trends in the unweighted comparison group were much closer to balanced for primary care visits than for specialty care visits.


\section{Conclusion}

The use of DD to recover causal effects in health policy research and other settings rests on the untestable assumption that intervention and comparison groups have parallel post-intervention trends in potential outcomes absent the intervention, an assumption that is frequently evaluated using information about pre-intervention outcome trends. When pre-intervention outcome trends are not parallel between intervention and comparison groups, unweighted DD estimators are likely to be biased and should be avoided. The results of this paper shows that it is possible to reduce bias by combining DD with weighting on pre-intervention outcome trends, and in certain contexts obtain unbiased estimates of causal effects.

The key distinction between the analysis presented here and other recent work that reached more pessimistic conclusions about combining matching with DD \supercite{daw2018matching} is that we examined DGPs where heterogeneity in individual-level outcome trends can be found within the intervention and comparison groups. When there is common support, or \textit{overlap}, between the intervention and comparison groups in the distribution of individual-level outcome trends, estimation of the microlevel DD after entropy balancing can greatly reduce the bias of estimates of the average treatment effect on the treated. We believe that such heterogeneity is likely to exist in many real-world settings of interest to health policy researchers. In these cases, our results provide a method for reducing bias in DD estimation even when pre-intervention group mean outcomes suggest a violation of the parallel trends assumption.

\begin{table}
    \centering
    \begin{tabular}{lcccc} 
          \multicolumn{3}{c}{}& \multicolumn{2}{c}{Entropy Balanced Comparison} \\ \cline{4-5}
          Measure & VBID & Comparison & Primary Care & Specialty Care \\ \hline
         Effective Sample Size & 39,570 & 36,588 & 35,072.4 & 34,736.8\\ 
         Baseline characteristics & & & & \\
         ~~Female (\%) & 54.2 & 55.2 & 54.2 & 54.2 \\ 
         ~~Age (Mean) & 77.8 & 77.1 & 77.8 & 77.8 \\ 
         ~~Dual Eligible (\%)  & 7.4 & 7.4 & 7.4 & 7.44 \\ 
         ~~Low-income Subsidy (\%) & 11.8 & 11.2 & 11.8 & 11.8 \\ 
         ~~Risk Score (Mean) & 1.80 & 1.58 & 1.80 & 1.80 \\ 
         ~~Disabled (\%)  & 15.4 & 14.8 & 15.4 & 15.4 \\ 
         ~~ESRD (\%) & 1.7 & 1.5 & 1.7 & 1.7 \\ 
         Baseline outcomes & & & & \\
         ~~Primary Care Visits (Mean) &4.9 & 4.4 & 4.6 & --- \\ 
         ~~Specialty Visits (Mean) & 10.4 & 8.4 & --- & 9.3 \\ \hline
    \end{tabular}
    \caption{Beneficiary characteristics, 2016, before and after entropy balancing.}
    \label{tab:descriptives}
\end{table}

\begin{table}
    \centering
    \begin{tabular}{lcc}
         Outcome & Unweighted & Entropy Balanced \\ \hline
         Primary care visits & 0.12 (0.08--0.17) & 0.16 (0.11--0.21) \\
         Specialty care visits & 0.05 (-0.04--0.14) & 0.14 (0.04--0.25)\\ \hline
    \end{tabular}
    \caption{Increase in the number of visits per beneficiary from difference-in-differences analyses, with and without entropy balancing.}
    \label{tab:results}
\end{table}

\printbibliography

\newpage
\section*{Appendix}
\subsection*{Weighted Identification of ATT in Observed Data}

In what follows, we define the observed and counterfactual outcome trends as derivatives for continuous time, which implicitly assumes such derivatives exist. However, the results do not require continuous time and the trends can instead be defined as differences between finite time points, and all results will hold. 

Recall that exposure groups are defined as $A = a \in \{0,1\}$, time is represented as $t \in \mathcal{T}$ and that there is an assumption that exposure is ultimately applied at $t_e>t_0$.  Effects are defined in the potential outcomes framework of Neyman/Rubin. We also make the usual Stable Unit Treatment Value assumption.\supercite{rubin2005causal} The following summary of previously defined notation may be helpful.

\begin{table}[h!]
    \centering
    \begin{tabular}{r|l}
        $Y_i(a,t)$ & Potential outcome under $A=a$ at time $T=t$ \\
        $Y_i(t)$ & Observed outcome at time $T=t$ \\
        $\tau_i(t) = Y_i(1,t) - Y_i(0,t)$ & Individual Effect at time $T=t$ \\
        $ATT(t) = E[\tau(t) | A = 1]$ & Average treatment effect on the treated \\ 
        $\delta_i(a,t) = \frac{\partial}{\partial t} Y_i(a,t)$ & Potential trend under $A=a$ at time $T=t$ \\
        $\delta_i(t) = \frac{\partial}{\partial t} Y_i(t) $ & Observed trend 
    \end{tabular}
    \caption{Summary of Notation}
    \label{tab:notationsummary}
\end{table}

As described previously, we assume the existence of a set of weights $w$ that demonstrate the following properties:

\begin{enumerate}[1)]
    \item \textit{Parallel Counterfactual Trends in the Post-Period}: $E[\delta(0,t) | A=1] = E[ w \delta(0,t) | A=0] \mbox{ for } t \geq t_e $.  
    
    \item \textit{Parallel Observed Trends in the Pre-Period:} $E[\delta(t) | A=1] = E[ w \delta(t) | A=0] \mbox{ for } t < t_e$
    
    \item \textit{Persistence of Similarity of Counterfactual Trends:} 
    
    $E[\delta(0,t) | A = 1] = E[w\delta(0,t) | A = 0] \mbox{ for } t < t_e \implies E[\delta(0,t) | A = 1] = E[w\delta(0,t) | A = 0] \mbox{ for } t \geq t_e$

\end{enumerate}
If it is possible to find weights that exhibit these properties, then it possible to identify the $ATT(t)$ from observed data with the following estimand, 
\begin{align*}
    DID(t) = E[Y(t) - Y(t_e) | A = 1] - E[w (Y(t) - Y(t_e)) | A = 0]
\end{align*}
The following claims will support the above assertion

\begin{description}
    \item[\textbf{Claim 1:}]  If Assumptions (2) and (3) hold, then Assumption (1) is satisfied, i.e., common observed pre-period trends and the persistence assumption, together, imply common post-period trends. 
    
    \textbf{Proof:}
    \begin{align*}
        E[\delta(t) | A = 1] &= E[w\delta(t) | A = 0] \mbox{ for all } t < t_e && \mbox{ (Assumption 2)} \\ 
        E[\delta(1,t) | A = 1] &= E[w\delta(0,t) | A = 0] \mbox{ for all } t < t_e && \mbox{ (Consistency)} \\ 
        E[\delta(0,t) | A = 1] &= E[w\delta(0,t) | A = 0] \mbox{ for all } t < t_e && (Y(1,t) = Y(0,t) \mbox{ for } t < t_e \implies \delta(1,t) = \delta(0,t)) \\
        E[\delta(0,t) | A = 1] &= E[w\delta(0,t) | A = 0] \mbox{ for all } t \ge t_e && \mbox{ (Assumption 3)}
    \end{align*}
    
    \item{\textbf{Claim 2:}} Assumption (1), along with suitable regularity conditions, implies that 
    \begin{align*}
        E[w (Y(0,t) - Y(0,t_e)) | A = 0] = E[Y(0,t) - Y(0,t_e) | A = 1] \mbox{ for } t \geq t_e
    \end{align*}
    \textbf{Proof:}
    \begin{align*}
          E[w (Y(0,t) - Y(0,t_e)) | A = 0] =& E\left[w \left(\int_{t_e}^t \delta(0,s) \partial s\right) | A = 0\right]  \\
        =& \int_{t_e}^t E\left[w \delta(0,s)  | A = 0\right] \partial s \\
        =& \int_{t_e}^t E\left[\delta(0,s)  | A = 1\right] \partial s \hspace{5em} \mbox{ (Assumption 1)} \\
        =&  E\left[ \left(\int_{t_e}^t\delta(0,s)\partial s\right)  | A = 1\right] \\
        =&  E\left[ Y(0,t) - Y(0,t_e)  | A = 1\right] \\
    \end{align*}
    
    \item{\textbf{Claim 3:}} Under Assumption (1) and Claim (2), the ATT is identified by the following 
    \begin{align*}
        DID(t) = E[Y(t) - Y(t_e) | A = 1] - E[w (Y(t) - Y(t_e)) | A = 0]
    \end{align*}

    \textbf{Proof: }. By consistency, it follows that
    \begin{align*}
        DID(t) &= E[Y(1,t) - Y(1,t_e) | A = 1] - E[w (Y(0,t) - Y(0,t_e)) | A = 0]\\ 
        &= E[Y(1,t) - Y(1,t_e) | A = 1] - E[Y(0,t) - Y(0,t_e) | A = 1] \hspace{3em} \mbox{(Claim (2))}\\ 
        &= E[Y(1,t) - Y(0,t_e) | A = 1] - E[Y(0,t) - Y(0,t_e) | A = 1] \hspace{3em} (Y(0,t_e) = Y(1,t_e))\\ 
        &= E[Y(1,t) - Y(0,t) | A = 1]
    \end{align*}

\end{description}

\subsection*{Entropy Balancing Details}

Entropy balancing addresses the problem of estimating weights by selecting weights that satisfy a series of constraints based on the balance of the observables between intervention groups. Our strategy is to develop additional constraints that the optimization must satisfy with respect to the trends of the observed outcomes, such that the resulting weights satisfy Assumption \ref{assumpt:observed}. 

First, we define the set of baseline weights $q_i = 1 / N_{0}$ where $N_{a}$ is the total number of observations for which $A=a$.  Next, we define a series of $m$ estimators $\hat d_i^m$ of the observed trend $\delta_i(t)$, or summary statistics of the trend, for each individual $i$ in the pre-period. (Two approaches to specifying $\hat d_i^m$ are discussed after we introduce the minimization problem that defines entropy balancing.)  Finally, we define a collection of $p$ variables $Z = \{Z_1, Z_2, \dots, Z_c, \dots, Z_p\}$ other than the pre-intervention trends that represent auxiliary information that may be important to balance in this procedure (e.g., including constraints that the observed covariates are ``balanced'').

The entropy balancing weights can then be defined as the solution of the following constrained minimization problem:
\begin{align*}
   &&& \min_{w} \sum_{i:A_i = 0} w_i \log\left(\frac{w_i}{q_i}\right) & \mbox{ subject to constraints that } \\
  1)&& & \sum_{i:A_i = 0} w_i h_j(\hat d_i^m) = E[h_j(d^m) | A = 1] & \mbox{ for each } j \mbox{ functions } h_j(\cdot) \mbox{ and estimator } m \\ 
  2)&& & \sum_{i:A_i = 0} w_i g_k(Z_{ic}) = E[g_k(Z_c) | A = 1] & \mbox{ for each } k \mbox{ function } g_k(\cdot) \mbox{ and each column } c \mbox{ of } Z \\ 
  3)&& & \sum_{i:A_i = 0} w_i = 1 \\
  4)&& & w_i > 0 & \mbox{ for all } i \mbox{ such that } A_i = 0 & 
\end{align*}


The functions $h_j(\cdot)$ and $g_k(\cdot)$ are typically chosen to represent moment conditions on the marginal distributions of the variables included in the balancing constraints. For instance, $\sum_{i:A_i=0} w_i Z_c = E[Z_c | A= 1]$ provides balance on the first of the variable $Z_c$ and $\sum_{i:A_i=0} w_i Z_c^2 = E[Z_c^2 | A= 1]$ would provide balance on the second moment. Additional functions $h_j(\cdot)$ could be specified to capture higher-order moments or covariances across time periods.

To apply entropy balancing to DD estimation, the most important variables to include in the balancing constraints are the estimators $\hat d_i^m $ of the pre-intervention outcome trends $\delta_i(t), t < t_e$; additional pre-intervention covariates may optionally be included in $Z_c$.  
Common practice in entropy balancing is to include balance constraints on the first two moments of the distributions that one wishes to balance ($h_1(z) = z$ and $h_2(z) = z^2$). The target values for each $E[g_k(Z_c) | A = 1]$ and each $E[h_j(d^m(t)) | A = 1]$ are calculated as sample averages in the intervention group.  \authorcite{hainmueller2012entropy} shows that the optimization above can be performed efficiently by first using the method of Lagrange multipliers to obtain a closed form solution for $w$ as a function of the remaining parameters and data and then solving the resulting dual problem using quasi-Newton methods.  Finally, weights are set to empirical weights for all observations in the exposed group, i.e..  $w_i = 1 / N_{1}$ for all $i$ such that $A_i = 1$.  

To incorporate pre-intervention trends into entropy balancing, we consider two approaches to specifying the trend estimators $ \hat d_i^m $. The first strategy is to nonparametrically estimate the slope between each time point, where if we consider a series of observed time points and define $m=\{0,1,\dots,M-1, M\}$ as an index of the observed time points such that $t_m < t_e$, i.e, $[t_0, t_1, \dots, t_M] < t_e$, we can define a series of non-parametric estimators of each individual's trend as 
\begin{align*}
    \hat d_i^m = \frac{Y_i(t_m) - Y_i(t_{m-1})}{t_m - t_{m-1}} \mbox{ for } m = 1, \dots, M
\end{align*}
In the language of time-series analysis, the series $\{ \hat d_i^m \}_{m=1}^M$ contains the first differences of $Y_i(t)$. 
If the observed trend is not ``noisy''---in the sense of being relatively free of white noise or other transitory error components---then we expect that this method will perform well and capture nonlinearities in the time-trend of the observed outcomes.

Unfortunately, when the time series of outcomes contains significant amounts of within-individual variability, this method is expected to have inadequate performance (as our algorithm may be balancing noise, introducing bias similar to that emphasized by \authorcite{daw2018matching}). An alternative approach that we explore is to parametrically model the relationship between the observed outcome and time using linear regression. Specifically, we approximate $E[Y_i(t)]$ as a function of time by estimating a regression of pre-intervention outcomes ($Y_i(t), t < t_e$) on a constant and a polynomial in time of order $M$:

\begin{align*}
Y_i(t) = \beta_{i0} + \sum_{m=1}^M \beta_{im} t^m + \varepsilon_{it}
\end{align*}

Under this model the individual trend in pre-intervention outcomes is captured through the $\beta_{im}$ coefficients for $m>0$., i.e., 
\begin{align*}
    \hat d_i^m = \hat \beta_{im} \mbox{ for } m = 1, \dots, M
\end{align*}
In this paper, we consider entropy balancing using parametric models with $M=1$ (linear trends) and $M=2$ (quadratic trends).  This parametric balancing may have the added benefit of smoothing over within-individual variability in outcome trends prior to balancing. However, we conjecture that misspecification of the parametric trends used to define entropy-balancing weights in the pre-intervention data may result in violation of Assumption 3 since the parametrically estimated pre-intervention trends would no longer predict the post-intervention trends. We explore this issue below in simulations where we use entropy balancing on linear pre-intervention trends when the true data generating process has quadratic trends.

\newpage

\subsection*{Extended Simulation Descriptions}

For all simulations, let $y_{it}$ denote the outcome of observation $i$ at time $t$, and let $A_{i}$ be an indicator of whether observation $i$ received the intervention. Assume $K_{pre}$ equally spaced time points prior to the intervention and $K_{post}$ time points after the intervention, with $K=K_{pre}+K_{post}$ total time points. We generate data from a multivariate normal distribution such that:

\begin{itemize}
 \item $\boldsymbol{y}_i =
        \begin{pmatrix}
            y_{i1} \\
            y_{i2} \\
            \vdots \\
            y_{iK} 
        \end{pmatrix} 
    \sim N\left( \boldsymbol{\mu}_i = 
        \begin{pmatrix}
            \mu_{i1} \\
            \mu_{i2} \\
            \vdots \\
            \mu_{iK} 
        \end{pmatrix}  ~~,~~ \boldsymbol{\Sigma} \right)$
 \item $\mu_{it} = E[y_{it} | A_{i}] = \beta_{0i} + \beta_{1i}t + \beta_{2i}t^2 + \tau A_{i} I(t > K_{pre}) $ 
    \item $(\boldsymbol{\beta}_i | A_i=a)  \sim N\left( \boldsymbol{\nu}_a ~~,~~ \boldsymbol{\Gamma}_a \right)$ are random effects that control the level of heterogeneity in outcome trends over time across individuals.  
\item $\boldsymbol{\Sigma}$ is an homoscedastic autoregressive covariance matrix of order $1$ with autoregressive parameter $\rho_y$ and variance given by $\sigma^2$.
\end{itemize}

\begin{table}[ht!]
    \centering
    \begin{tabular}{|l|c|c|c|} \cline{3-4}
        \multicolumn{2}{c}{} & \multicolumn{2}{|c|}{\bfseries{Model parameters}} \\ \hline
        \multicolumn{1}{|c|}{\bfseries{Simulation}}  & \bfseries{Time points} & Time Effects  & Heterogeneity  \\ \hline 
        \multirow{9}{*}{$\begin{array}{l} \textbf{Scenario 1}\\ \\ \text{Linear time effect with} \\ \text{overlapping individual heterogeneity} \end{array} $}  & \multirow{9}{*}{$\begin{array}{l} K_{pre}= 4\\ K_{post}=1 \end{array} $} & & \\
        &    &  
        \multirow{3}{*}{
            $\boldsymbol{\nu}_0=\begin{pmatrix}
                0 \\
                0 \\
                0 
            \end{pmatrix}$} & \multirow{3}{*}{
        $\boldsymbol{\Gamma}_0 = \begin{pmatrix}
                0 & 0 & 0 \\
                 & 0.2^2 & 0 \\
                 &  & 0
            \end{pmatrix}$ } \\
           &    &  & \\ 
           &  & & \\
           &  & & \\
                    &  &       \multirow{3}{*}{
            $\boldsymbol{\nu}_1=\begin{pmatrix}
                1 \\
                -0.2 \\
                0 
            \end{pmatrix}$ } & \multirow{3}{*}{
        $\boldsymbol{\Gamma}_1 = \begin{pmatrix}
                0 & 0 & 0 \\
                 & 0.1^2 & 0 \\
                 &  & 0
            \end{pmatrix}$} \\
                     &  & &   \\
                     &  & & \\
                    &  & & \\
           \hline
         \multirow{9}{*}{$\begin{array}{l} \textbf{Scenario 2}\\ \\ \text{Linear time effect with} \\ \text{non-overlapping individual heterogeneity} \end{array} $} &  \multirow{9}{*}{$\begin{array}{l} K_{pre}= 4\\ K_{post}=1 \end{array} $} & & \\
          &    & \multirow{3}{*}{
            $\boldsymbol{\nu}_0=\begin{pmatrix}
                0 \\
                -0.2 \\
                0 
            \end{pmatrix}$}  
        &   \multirow{3}{*}{$\boldsymbol{\Gamma}_0 = \begin{pmatrix}
                0 & 0 & 0 \\
                 & 0 & 0 \\
                 &  & 0
            \end{pmatrix}$ }\\
           &    &  &  \\ 
           &  & & \\
           &  & & \\
                    &  &       \multirow{3}{*}{
            $\boldsymbol{\nu}_1=\begin{pmatrix}
                1 \\
                0 \\
                0 
            \end{pmatrix}$ }  
        &   \multirow{3}{*}{$\boldsymbol{\Gamma}_1 = \begin{pmatrix}
                0 & 0 & 0 \\
                 & 0 & 0 \\
                 &  & 0
            \end{pmatrix}$}\\
                     &  & &  \\
                     &  & & \\
                     &  & & \\
           \hline
        \multirow{9}{*}{$\begin{array}{l} \textbf{Scenario 3}\\ \\ \text{Quadratic time effect with} \\ \text{individual heterogeneity} \end{array} $} &  \multirow{9}{*}{$\begin{array}{l} K_{pre}= 4\\ K_{post}=1 \end{array} $} & & \\
        &  & \multirow{3}{*}{
            $\boldsymbol{\nu}_0=\begin{pmatrix}
                0 \\
                0 \\
                0 
            \end{pmatrix}$} &  \multirow{3}{*}{$\boldsymbol{\Gamma}_0 = \begin{pmatrix}
                1^2 & 0.1 & -0.04 \\
                 & 0.2^2 & -0.0075 \\
                 &  & 0.05^2
            \end{pmatrix}$ }\\

           &    &  & \\ 
           &    &  & \\
           &    &  & \\ 
                    &  &       \multirow{3}{*}{
            $\boldsymbol{\nu}_1=\begin{pmatrix}
                1 \\
                -0.2 \\
                0.05 
            \end{pmatrix}$}  
        &  \multirow{3}{*}{$\boldsymbol{\Gamma}_1 =\begin{pmatrix}
                1^2 & 0.05 & -0.02 \\
                 & 0.1^2 & -0.001875 \\
                 &  & 0.025^2
            \end{pmatrix}$}\\
                     &  & &   \\
                     &  & & \\
                  & &  & \\
           \hline
    \end{tabular}
    \caption{Simulation Parameters. Unless noted elsewhere, $N_0=1000$, $N_1=500$, $\tau=0$, $\sigma^2=1$, and $\rho_y \in \{ 0, 0.1,0.3,0.5,0.7,0.9,0.99  \}$. $\nu \equiv (\beta_0, \beta_1, \beta_2)$ }
    \label{tab:simulation1}
\end{table}

\subsection*{Reliability of Pre-Intervention Trend Estimation Due to Variance of Error Term}

In the text, our simulations showed a general pattern of reduced bias as the residual autocorrelation increased. This is due to the fact that with higher residual autocorrelation, we are better able to distinguish individual-level outcome trends from random variation. To further explore the importance of reliability, we modified the DGP from Scenario 1 by holding the residual autocorrelation fixed and changing other simulation parameters to manipulate reliability. 

\begin{figure}
    \centering
    \includegraphics[width=\textwidth]{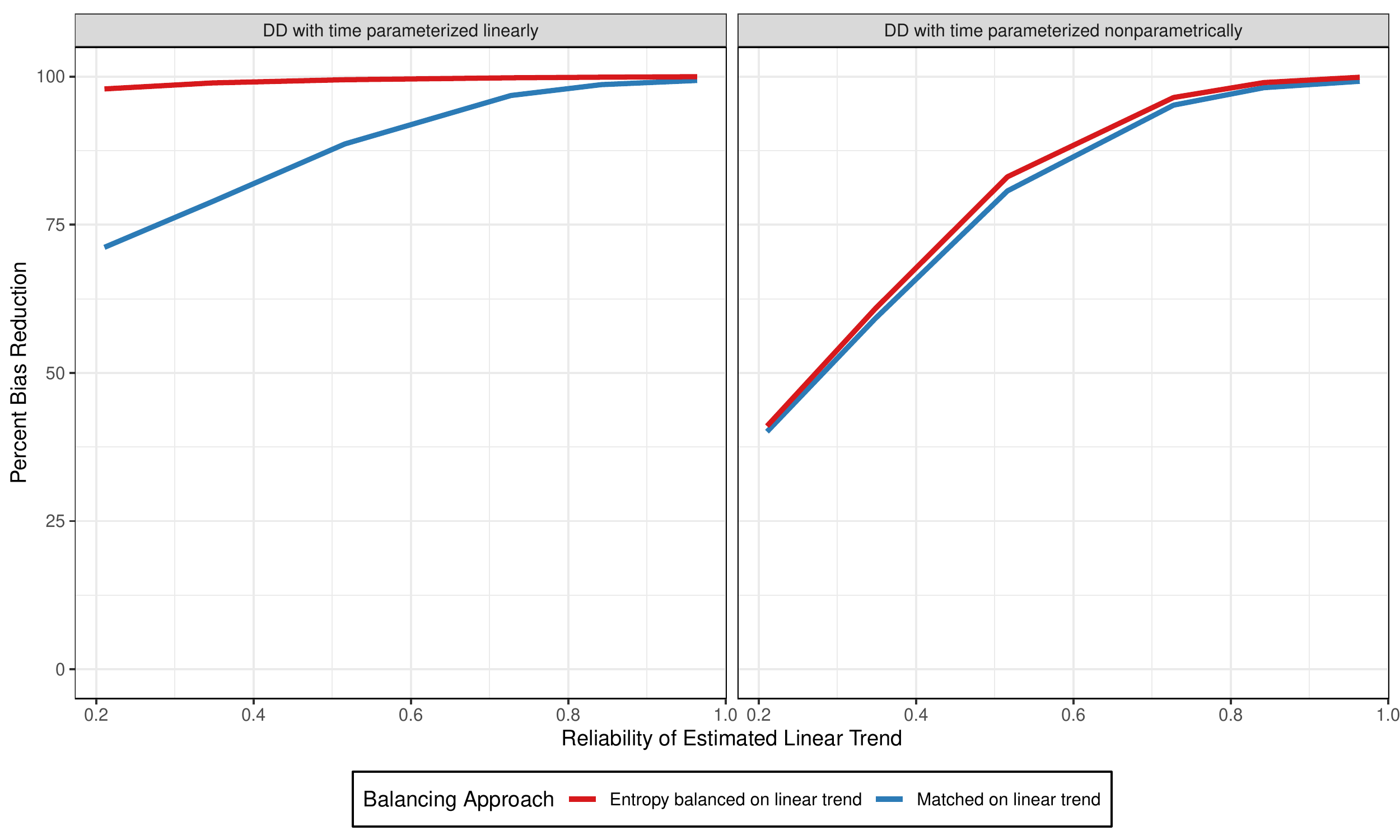}
    \caption{Percent bias reduction as a function of the reliability of the estimated linear trends from the simulation study when group counterfactual trends are different with overlap in the individual-level trends. The left panel provides the results for DD with time parameterized linearly, and the right panel provides the results for DD with time parameterized nonparametrically.}
    \label{fig:fig4}
\end{figure}

To explore a different mechanism that could affect reliability, we reproduced the simulation when the group counterfactual trends are different with overlap, but fixed the residual autocorrelation to be $0.5$ and varied the residual variance from $0.01$ to $1$. For each value of the residual variance, we calculated the reliability of the estimated linear trend by comparing the between-individual variance in the linear trends among the comparison group ($\nu_0=0.2^2$) with the within-individual variance in the estimated linear trends. Figure \ref{fig:fig4} provides the percent bias reduction comparing each estimator to the usual DD for this scenario, varying the reliability of the estimated linear trends. We note that the bias reduction increases as the reliability increases. Overall, the bias reduction results are very similar to those in Figure \ref{fig:fig1} despite the fact that differences in reliability are driven by a different statistical mechanism.


\subsection*{Nonlinear counterfactual trends}

In this section, we explore the performance of balancing pre-intervention outcome trends when there are nonlinearities in the counterfactual trends. We allowed for slight nonlinearity by including a quadratic function of time, which varies across individuals. We excluded matching from these results. Further, since the counterfactual trends are nonlinear, we included the entropy balancing approach that nonparametrically balances the pre-intervention trends. 

\begin{figure}
    \centering
    \includegraphics[width=\textwidth]{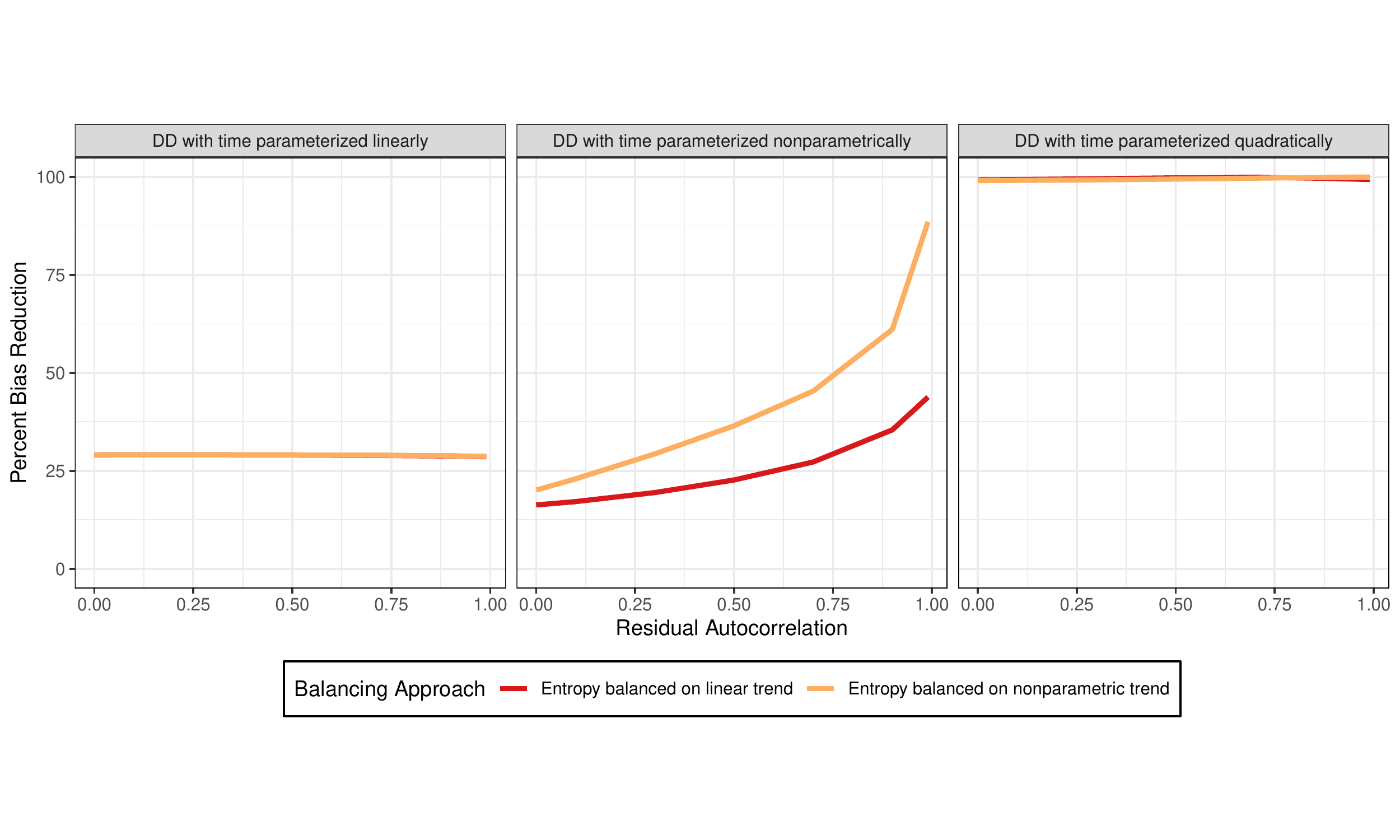}
    \caption{Percent bias reduction from the simulation study with nonlinear counterfactual trends. The left panel provides the results for DD with time parameterized linearly, the middle panel provides the results for DD with time parameterized nonparametrically, and the right panel provides the results for DD with time parameterized quadratically.}
    \label{fig:fig3}
\end{figure}

Figure \ref{fig:fig3} provides the percent bias reduction comparing each estimator to the usual DD for this scenario, varying the residual autocorrelation. A small bias reduction is observed using either balancing approach when the DD model includes times linearly (left panel). Moving to the DD model that includes time nonparametrically (middle panel), we note that both entropy balancing approaches reduce bias compared to the unadjusted difference-in-difference model. However, entropy balancing the nonparametric trends has greater bias reduction than entropy balancing the linear trends, and its bias tends towards 0 as the autocorrelation increases. This highlights the benefits of using nonparametric time simultaneously in entropy balancing and the DD model. While this specification does not produce unbiased estimates for all values of the autocorrelation in this simulation, more generally speaking, we expect the amount of bias reduction from the nonparametric version of this approach to be related to the reliability of the estimated trends, or in other words, our ability to distinguish the individual trends from random noise \supercite{steiner2011importance}. This will be achieved as the autocorrelation increases, as the residual error decreases, and as the within-group variance of the time trends increase, among others. The last panel of Figure \ref{fig:fig3} provides the bias when specifying a DD model with a quadratic time trend. Here, we note that both entropy balancing approaches are approximately unbiased, despite entropy balancing of the linear time trends being a misspecification. 

\newpage
\subsection*{Additional Details for MA VBID Analysis}
For ease of presentation, the analysis presented in this paper departs from the analyses published in \authorcite{eibner2018first} in three important respects. First, to abstract from attrition and missing data issues, analyses were restricted to beneficiaries with complete data from 2014 to 2017. Second, for simplicity of reporting, we report linear DD models estimated using generalized estimating equations (GEE) with an unstructured working correlation structure to account for within-beneficiary serial correlation. \authorcite{eibner2018first} used nonlinear models to account for the skewed distribution of our count outcome variables. Finally, along with non-parametric outcome trends, we included the full set of baseline characteristics from Table \ref{tab:descriptives} in the entropy balancing algorithm.


\end{document}